\documentclass[a4paper, aps, pra, superscriptaddress, twocolumn]{revtex4-1}
\usepackage[utf8]{inputenc}
\usepackage[english]{babel}
\usepackage{graphicx}

\begin{document}

\title{Measuring the Orbital Angular Momentum of Electron Beams}

\date{\today}

\author{Giulio Guzzinati}
\affiliation{EMAT, University of Antwerp, Groenenborgerlaan 171, 2020 Antwerp, Belgium}

\author{Laura Clark}
\affiliation{EMAT, University of Antwerp, Groenenborgerlaan 171, 2020 Antwerp, Belgium}

\author{Armand Béché}
\affiliation{EMAT, University of Antwerp, Groenenborgerlaan 171, 2020 Antwerp, Belgium}

\author{Jo Verbeeck}
\affiliation{EMAT, University of Antwerp, Groenenborgerlaan 171, 2020 Antwerp, Belgium}

\begin{abstract}
The recent demonstration of electron vortex beams has opened up the new possibility of studying orbital angular momentum (OAM) in the interaction between electron beams and matter.
To this aim, methods to analyze the OAM of an electron beam are fundamentally important and a necessary next step.
We demonstrate the measurement of electron beam OAM through a variety of techniques.
The use of forked holographic masks, diffraction from geometric apertures,
diffraction from a knife-edge and the application of an astigmatic lens are all experimentally demonstrated.
The viability and limitations of each are discussed with supporting numerical simulations.
\end{abstract}

\maketitle

Following the discovery that particles also possess wave properties, electrons have been employed as
a powerful tool to study the microscopic and fundamental properties of matter, being employed in a 
wide range of diffraction techniques and spectroscopies.
These techniques rely on the determination of the energy or linear momentum of electrons.
The recent prediction \cite{Bliokh2007} and realization \cite{Uchida2010,Verbeeck2010,McMorran2011}
of electron vortex beams, possessing orbital angular momentum(OAM), has opened up the possibility
of studying the role of OAM in the interaction between electron beams and matter. This brings new
possibilities to study magnetism, nanoparticle manipulation and rotational friction in the TEM 
\cite{Verbeeck2010,Verbeeck2013}.

Electron vortex beams are paraxial beams with helical wavefronts of the form $A(r,z) \exp(im\phi) $,
where $m$ is the topological charge. Electrons in such states possess an OAM of $m \hbar$.
It is worth noting that the direct proportionality between topological charge and OAM is only
verified as long as the intensity distribution is cylindrically symmetric around the phase dislocation 
\cite{ONeil2002}. For simplicity  we will only consider this case as the
generalised case can be studied as superposition of such states.

In order to explore the role of angular momentum in beam-sample interaction it is important to have both
control over the OAM of the incident beam and the ability to quantify the OAM of the outgoing wave.

This problem has been studied extensively for optical vortices, and many solutions have
been devised. A simple interference with a reference wave creates unique patterns that allow
direct determination of the OAM. Other methods include using multipoint interferometers
\cite{Berkhout2008}, geometric transformations by phase manipulation
\cite{Berkhout2010, Lavery2011b},
and the use of multiple interferometers in a cascade setup
\cite{Leach2002}.

While significant effort has been put into the generation of electron vortex beams \cite{Uchida2010,Verbeeck2010,McMorran2011,Verbeeck2011a,Saitoh2012a,Clark2013,Beche2013},
little progress has been made in measuring the OAM and the lower flexibility of existing electron-optical components prevents the application of the techniques mentioned above.
In this Letter we begin bridging this gap, demonstrating different methods to detect and quantify the
OAM of an electron vortex beam.

On the quantum mechanical level, the OAM of a paraxial wave can be calculated integrating
over the whole plane the orbital angular momentum density,
defined as $\mathbf{r}\times\mathbf{p}_\varphi$ where ${\mathbf{r}}$ is the position operator
and $\mathbf{p}_\varphi$ is the azimuthal component of the linear momentum density operator.
An ideal method should be able to measure this quantity independently of the radial component of $\mathbf{p}$, and without hypothesis on the shape of the wave. This can be done e.g. conformally mapping $ \mathbf{p}_\varphi $ and $ \mathbf{p}_r $ into $ \mathbf{p}_x $ and $ \mathbf{p}_y $ through ad-hoc phase-plates \cite{Berkhout2010}, but the limitations of phase-plate technology for electrons prevent the use of this method.
As we will see, all the methods presented here fall short of this strict requirement.

\begin{figure}[htbp]
    \centering
    \includegraphics[width=\columnwidth]{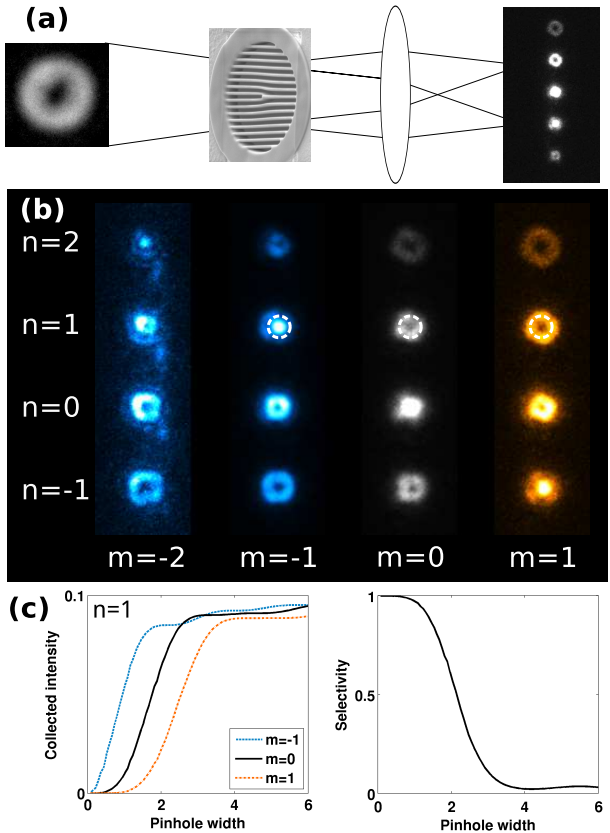}
    \caption{(color online) OAM measurement with a fork grating.
	     (a) Schematic of the setup.
	     (b) Experimental images of the diffraction pattern produced by the fork
	         for the values of incident OAM $m=\{-2,-1,0,1\}$.
	         Each column represents a recorded pattern, 
	         the diffraction order of each beam is indicated on the left.
	     (c) Simulation of a signal collected by a pinhole positioned on the first diffracted order,
	         as a function of the pinhole width and for different illuminating beams, and selectivity
		 of the pinhole for dichroic signal defined as $ {(I_{-1}-I_{1})}/{(I_{-1}+I_{1})} $.
		 The pinhole size is normalized to the FWHM of an $m=0$ beam.
    \label{fig:fork_sa}}
\end{figure}
Previously measurement of the OAM of electron beams has been performed through the computer
generated holograms (CGH) that have been used for the production of vortex beams \cite{Verbeeck2010,Saitoh2013}.
These CGH are
gratings with a
dislocation, calculated numerically interfering a vortex
beam with a reference plane wave. An incoming plane wave is diffracted by the CGH into a 1D vortex array. 
The OAM of each diffraction order is $m = n \ell$ where $\ell$ is the dislocation order and $n$ the
diffraction order. The intensity of the various spots depends on the bar-with/slit-width ratio that determines the single-slit-envelope of the intensities \cite{Born1999}.

Illuminating the mask with a vortex beam of OAM $m_i$ (see figure \ref{fig:fork_sa}a) the
OAM in the diffracted spots changes to $m = m_i+n \ell$, while the relative intensities of the different
diffraction spots are, to a good approximation, unaltered \cite{Topuzoski2011,Saitoh2013}. The phase
discontinuity is not present in the $m=0$ beam, and so it does not acquire the characteristic doughnut intensity
profile \cite{Janicijevic2008}.

We experimentally verified this by placing an $\ell =1$ fork aperture in the illumination system of the X-Ant-EM microscope 
\footnote{The X-Ant-EM is an FEI Titan$^3$ installed at the University of Antwerp}
operating at 200\,kV,
and using the resulting vortex beams to illuminate a second $\ell =1$ fork aperture placed in the projection system. Switching the magnetic-lens system to diffraction mode projects the far-field diffraction of the aperture
onto a CCD camera. The OAM of the input beam can be deduced observing which diffraction order does not possess a doughnut
intensity profile, thus satisfying $m_i + n\ell = 0 $, as illustrated in figure \ref{fig:fork_sa}b.

As vortex beams
possess a central intensity minimum whose width scales with $\sqrt{|m|}$, a pinhole placed in the position
of the diffracted beam can discriminate between a vortex or a non-vortex beam, analogous to the use of a single-mode fiber in light optics \cite{Mair2001}.

This method is inefficient as the absorption from the mask and the further subdivision of intensity between
different beams leaves only ~10\% of the initial intensity in the first order diffracted beam, even
less in higher orders. Additionally the discrimination is more accurate for a smaller pinhole, with
the result that most of the beam's intensity is lost.

In order to estimate the discrimination efficiency of this technique we simulated the
intensity collected by a pinhole. We simulated the diffraction pattern produced
by an ideal fork mask when illuminated with vortex beams with $m=\{-1,0,1\}$,
then we integrated the intensity scattered within a circular aperture centered on
the $n=1$ diffraction order, and plotted the intensity as a function of the radius of the aperture for the
different values of incident OAM. The intensity is normalized to the incident intensity
in the single $m=\{-1,0,1\}$
beam, and the pinhole size is normalized to the FWHM width of an $m=0$ beam. The selectivity, defined as
${(I_{-1}-I_{1})}/{(I_{-1}+I_{1})}$ where $I_n$ is the intensity collected from the incoming component with $m=n$, is also shown.

We found that if a high selectivity is required, the signal is extremely low. With a normalized pinhole
diameter of 1, the selectivity is $\sim 0.97$ and the intensity as low as $5\%$ of the incident intensity.
With a pinhole diameter of 2.5 the collected intensity is increased to $8.5\%$ but the selectivity decreases already to $\sim 0.27$.

For higher order beams the detection efficiency is even lower due to the weaker intensities of the Bragg spots.

It should be noted that the applicability of this simulation is limited as the radial shape of the diffracted beams,
and therefore the detection efficiency, depends on the radial intensity distribution of the beam incident
on the mask. In the extreme case, where the radial distribution is entirely unknown, the OAM selectivity
is achieved only in the very center of the diffracted spot.

\begin{figure}
    \centering
    \includegraphics[width=\columnwidth]{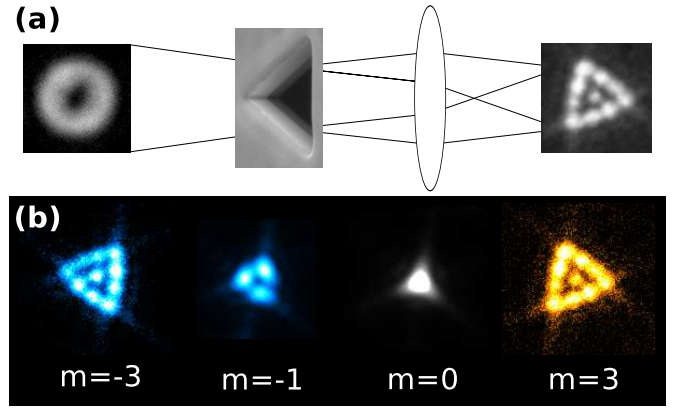}
    \caption{(color online) OAM measurement with a triangular aperture.
             (a) Schematic representation of the setup.
             (b) Experimental images, the absolute OAM value can be deduced counting the spots
	         in the pattern, as the triangle's side will have $|m|+1$ lobes.
    \label{fig:triangles}}
\end{figure}
The previous case shows how the phase singularity of a vortex beam determines the diffraction pattern it produces. One might wonder whether replacing the CGH with a different binary aperture would allow  identification of
the OAM while conserving a greater fraction of the incoming intensity. 
Indeed using geometrically shaped apertures can produce characteristic diffraction patterns
that allow the identification of the topological charge of the incident beam 
\cite{Guo2009,Hickmann2010,Liu2011a,Liu2011b,Liu2011c}.
Among the various examples the triangular aperture is particularly interesting due to its simple analysis.

The diffraction of a vortex beam by a triangular aperture produces a triangular lattice in the
far-field which is determined by the input topological charge.
The origin of this pattern can be understood recalling that the diffraction of a wave
by an aperture is formed by the interference between the edge waves.
The extra phase in an incident vortex beam shifts the edge waves,
forming a triangular pattern. The magnitude of the shift and thus the size of
the pattern is determined by the value of $|m|$. The handedness of the OAM relates the
orientations of pattern and aperture \cite{Hickmann2010}.

This method has been shown to hold also for vortex beams with non-integer topological charge, and the
rotation the pattern acquires upon changing the sign of the OAM has been linked to the Gouy phase \cite{Mourka2011}.
Therefore recording the diffraction pattern and analyzing arrangement and number of intensity maxima
allows retrieving both value and sign of the OAM \cite{Hickmann2010,Mourka2011}.

We verified this by placing a triangular aperture in a Philips CM30 TEM at 300\,kV. The vortex beams were created by a forked
hologram in the illumination system of the microscope and used to
illuminate the triangle, recording the diffraction pattern with a CCD camera
(see figure \ref{fig:triangles}a).

The resulting pattern shows the expected characteristics as shown in figure \ref{fig:triangles}b.
The number of maxima on the edge of the triangle scales as $|m|+1$ and the direction reverses upon changing the sign,
allowing easy identification of the OAM. The first limitation of this approach lies in the fact that
the analysis is fundamentally more complicated than simple signal counting as in the previous case.
Furthermore the analysis of such a pattern is simple only if the vortex beam is an OAM eigenstate.
A superposition of states produces diffraction patterns that deviate from the triangular lattice pattern
and are harder to interpret. 
An incoherent superposition of two vortex states generates a pattern consisting of the sum of the two
different patterns. When such a superposition is formed by modes of different $|m|$ the features of the
lower order mode tend to be more prominent, as the intensity is concentrated on a smaller area.

\begin{figure}
    \centering
    \includegraphics[width=\columnwidth]{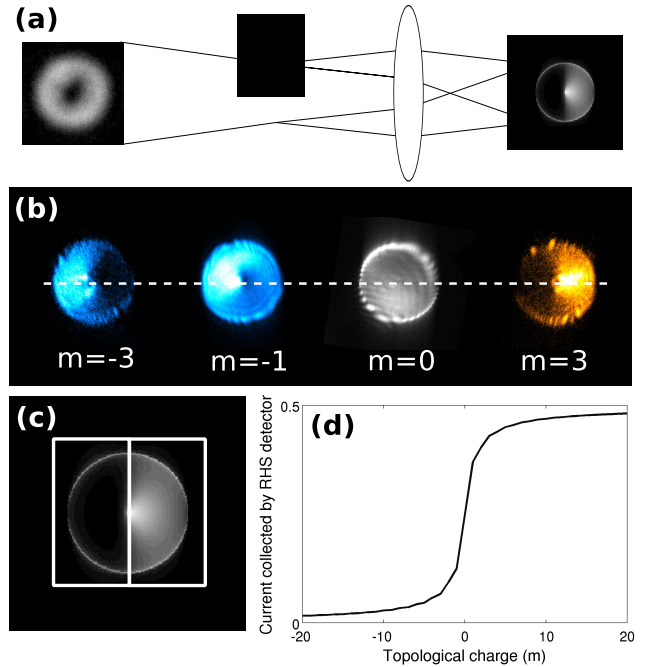}
    \caption{(color online) OAM measurement with a knife-edge.
	     (a) Schematic of the setup.
	     (b) Experimental images, the dashed line indicates the direction of the knife edge.
	     (c) Schematic representation of a detector divided in two parts with the dividing line
	         orthogonal to the knife-edge, and on which the beam has been centered.
             (d) The fraction of he total current that will reach
	         the right-hand-side of the detector as a function of the OAM.
    \label{fig:gouy}}
\end{figure}
Most of the techniques based on geometrical apertures produce a pattern that needs to be recorded and analyzed
in order to obtain the OAM. However a knife edge generates a diffraction pattern that lends itself to
the development of a counting-based technique.

While we have already shown that a knife edge can be used to reveal the handedness of an electron vortex beam
\cite{Guzzinati2013} the possibility for detecting the value of the OAM has only indirectly been explored \cite{Schattschneider2012}. 

If we block half of a vortex beam with a knife edge at the waist, thus obtaining a C-shaped beam,
we can observe that upon propagation the beam undergoes a deformation of the intensity pattern and 
a characteristic rotation whose direction depends on the sign of the angular momentum \cite{Arlt2003, Guzzinati2013}.
In the far-field we observe that for opposite values of OAM the patterns are rotated by
$\pi$\,radians with respect to each other, and possess an asymmetric intensity distribution.
Another way to interpret this phenomenon is that while the spiraling current-density of the vortex mode
possesses an average zero value of transverse momentum, blocking half of the beams breaks this symmetry,
and the resulting C-shaped beam has a non zero value of transverse momentum, leading to a shift in the
diffraction pattern.

We verified this experimentally in a Philips CM30 TEM. We selected a single vortex beam generated by the fork mask using a second aperture, then blocked half of this beam with the knife-edge (see figure \ref{fig:gouy}a).
The resulting patterns, shown in figure \ref{fig:gouy}b,
present the expected asymmetry and mirror symmetry upon changing the sign of the OAM.
Additionally the asymmetry appears stronger for higher OAM.

In order to explore the feasibility of this method, we performed numerical
simulations studying the link between the value of OAM and the asymmetry in the diffraction pattern.
For this we supposed a knife-edge blocking half of a vortex beam at its waist, then centering the
resulting far field pattern on a detector. We imagined this detector as divided in two parts with independent signal output of the impinging current,
as in figure \ref{fig:gouy}c. We then calculated the fraction of the intensity collected by
the right-hand-side of the detector, shown in figure \ref{fig:gouy}d.
It was found that this signal depends on the OAM in a nonlinear way, but appears to saturate at
a maximum value of half the incident intensity (half of the intensity is blocked by the knife edge). 

This shows that this method can only be applied to low values of OAM with reasonable accuracy.
Moreover, if this method is applied to analyze a superposition of states
the non-linearity makes it impossible to uniquely obtain the average OAM or
the relative weight of each mode.

However if the wave is known to be an incoherent superposition of $m=\{ - m_0 , 0 , m_0 \}$
a direct proportionality between signal and $m$ can be established,
enabling OAM measurement.

\begin{figure}
    \centering
    \includegraphics[width=\columnwidth]{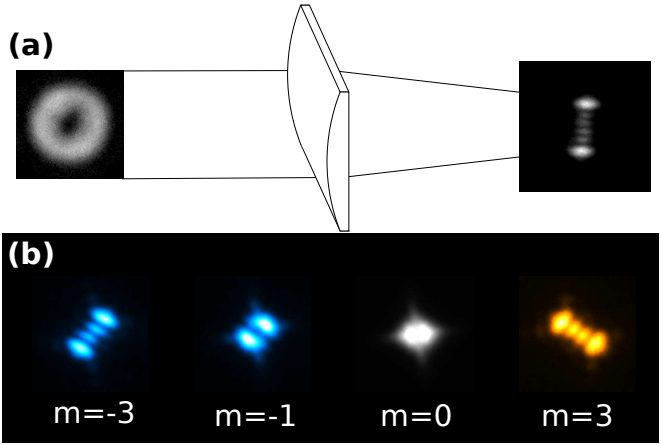}
    \caption{(color online) OAM measurement by astigmatic phase.
             (a) Schematic of the setup
	     (b) Experimental results.
    \label{fig:astigmatism}}
\end{figure}

While the above methods all employ binary diffraction 
techniques, in the optical case, the phase can be directly
manipulated to reveal the OAM \cite{Berkhout2010}.
Equivalent phase manipulation techniques within the TEM,
are currently not flexible enough to enable a true OAM decomposition;
while TEM phase manipulation has been demonstrated in the production of electron vortices \cite{Petersen2013, Clark2013}.
There is however a phase manipulation method of simple experimental realization that allows simple measurement of the topological charge, a method based on the mode-conversion process also used to produce vortex beams.

Typically a higher-order Hermite-Gaussian mode is converted, applying successive 
astigmatic phase shifts, into a higher-order Laguerre-Gaussian beam which carries a phase vortex.
This is achieved in light optics with cylindrical lenses, or in electron optics using the 
electron-optic stigmators \cite{Allen1992, Schattschneider2012a}.
The order of an LG-like vortex mode can be measured by reversing this process – applying a quadratic
phase-plate divides the doughnut intensity profile, into a number of linearly arranged intensity lobes,
where the number of lobes is equal to $|m|+1$. The orientation of the pattern with respect
to the phase-plate (angled at $\pi/4$ ), reveals the sign of $m$ \cite{Vaity2012, Vaity2013}.
As demonstrated in figure \ref{fig:astigmatism}b, the experimental results neatly follow these predictions.
This method is particularly easy to employ within the TEM, requiring the manual adjustment
of only one parameter, which is freely tunable on any electron microscope. Indeed this technique can be
an ideal way to confirm the vortex beam order during the preparation of a more complex experimental set-up
and then readjusted to an astigmatism-free condition. However
impure modes would lead to overlapping of the intensity lobes, so this technique only works for pure vortex states, presenting in this the same limitations as the triangular aperture.
Furthermore the characteristic pattern can only be observed close to the beam's waist.

We have presented and demonstrated several methods for the measurement of OAM in electron beams. Two of these demonstrations clarified the details and generalized the scope of the methods shown in earlier publications \cite{Verbeeck2010,Saitoh2013,Guzzinati2013} while two additional methods (the triangular aperture and the astigmatic phase) were demonstrated in TEM use for the first time, introducing additional flexibility in this newly developing field.

The methods employing the triangular aperture and the astigmatic phase allow the measurement of any order of topological charge, but require the characteristic pattern to be recorded and analyzed.
Alternatively the knife-edge and the fork mask are more suitable for analysis of low order vortex beams,
but potentially allow the measurement to be reduced to a simple counting
which could be automated. However the high versatility in this respect of the fork mask comes at
the expense of a very low detection efficiency, while if the above mentioned restrictions on the values to be measured can be imposed, the knife-edge grants a better efficiency.

The applicability of these methods is restricted to eigenstates of OAM or in some cases to incoherent superpositions of these states, and does not translate well to arbitrary beams, where the outcome in general depends not on the OAM alone but also on the exact form of the beam \cite{Jesus-Silva2012}.
While this sets a target for future development of the detection methods,
the applications can already benefit from these results.
A variety of phenomena can already be studied within these restrictions such as
the EMCD effect or the generation of vortex beams by magnetic monopoles \cite{Beche2013}.
We believe that the availability of methods to detecting the OAM will lead to new and interesting developments,
as the role of OAM is considered in phenomena such as diffraction \cite{Takahashi2013} or elastic propagation of
electron beams through matter \cite{Loffler2012,Lubk2013a}.

\begin{acknowledgments}
The authors acknowledge funding from the European Research Council
under the 7th Framework Program (FP7), ERC Starting Grant no. 278510 VORTEX.
and the European Research Council under the 7th Framework Program (FP7),
ERC Grant no. 246791-COUNTATOMS.
Financial support from the European Union under the Framework 7 program under a contract
for an Integrated Infrastructure Initiative (Reference No. 312483 ESTEEM2) is also gratefully acknowledged.
\end{acknowledgments}

\bibliographystyle{apsrev4-1}
\bibliography{biblio}

\end{document}